\documentclass{article}

\usepackage{arxiv}

\usepackage[utf8]{inputenc} 
\usepackage[T1]{fontenc}    
\usepackage{hyperref}       
\usepackage{url}            
\usepackage{booktabs}       
\usepackage{amsfonts}       
\usepackage{nicefrac}       
\usepackage{microtype}      
\usepackage{lipsum}		
\usepackage{graphicx}
\usepackage{doi}
\usepackage{siunitx}

\title{Monolithically 3D nano-printed mm-scale lens actuator for dynamic focus control in optical systems}

\author{ \href{https://orcid.org/0000-0003-4325-5129}{\includegraphics[scale=0.06]{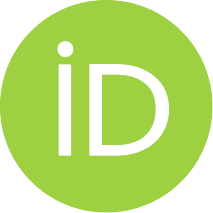}\hspace{1mm}Florian~Lux}\thanks{\href{mailto:florian.lux@uni-freiburg.de}{florian.lux@uni-freiburg.de}}\hspace{4pt} \\
		Microsystems for Biomedical Imaging Group\\
	Department of Microsystems Engineering (IMTEK)\\
	University of Freiburg\\
	79110 Freiburg im Breisgau, Germany\\
	 \\
	\And
	\href{https://orcid.org/0000-0003-2347-958X}{\includegraphics[scale=0.06]{orcid.pdf}\hspace{1mm}Aybuke~Calikoglu} \\
	Laboratory for Micro-optics\\
        Department of Microsystems Engineering (IMTEK)\\
	University of Freiburg\\
	79110 Freiburg im Breisgau, Germany\\
	 \\
        \And
	\href{https://orcid.org/0000-0002-6280-8465}{\includegraphics[scale=0.06]{orcid.pdf}\hspace{1mm}\c{C}a\u{g}lar~Ataman} \\
	Microsystems for Biomedical Imaging Group\\
	Department of Microsystems Engineering (IMTEK)\\
	University of Freiburg\\
	79110 Freiburg im Breisgau, Germany\\
    }

\renewcommand{\shorttitle}{Monolithically 3D nano-printed mm-scale lens actuator for dynamic focus control in optical systems}

\hypersetup{
pdftitle={Monolithically 3D nano-printed mm-scale lens actuator for dynamic focus control in optical systems},
pdfsubject={q-bio.NC, q-bio.QM},
pdfauthor={Florian Lux, Aybuke Calikoglu, Caglar Ataman},
pdfkeywords={MEMS scanner, two-photon polymerization, 3D nano-printing, dynamic focus control, electromagnetic actuation},
}

\begin{document}
\maketitle

\begin{abstract}
Three-dimensional (3D) nano-printing via two-photon polymerization offers unparalleled design flexibility and precision, thereby enabling rapid prototyping of advanced micro-optical elements and systems, including hybrid achromats, diffractive and flat optics, millimeter-sized lenses, fiber-optic sensor heads, and photonic waveguides. These elements have found important applications in endomicroscopy and biomedical imaging. The potential of this versatile tool for monolithic manufacturing of dynamic micro-opto-electro-mechanical systems (MOEMS), however, has not yet been sufficiently explored. This work introduces a 3D nano-printed lens actuator with a large optical aperture, optimized for remote focusing in miniaturized imaging systems. The device integrates ortho-planar linear motion springs, a self-aligned sintered micro-magnet, and a monolithic lens, actuated by dual micro-coils for uniaxial motion. The use of 3D nano-printing allows complete design freedom for the integrated optical lens, while the monolithic fabrication ensures inherent alignment of the lens with the mechanical elements. With a lens diameter of \SI{1.4}{\milli\meter} and a compact footprint of \SI{5.74}{\milli\meter}, it achieves high mechanical robustness at resonant frequencies exceeding \SI{300}{\hertz} while still providing large a displacement range of \SI{200}{\micro\meter} (\textpm\,\SI{100}{\micro\meter}). A comprehensive analysis of optical and mechanical performance, including the effects of coil temperature and polymer viscoelasticity, demonstrates its advantages over conventional MEMS actuators, showcasing its potential for next-generation imaging applications. 
\end{abstract}


\section{Introduction}
3D nano-printing via two-photon polymerization, also called direct laser writing, offers unprecedented design freedom and precision for micro-optical elements and systems. Following first demonstrations of single microlenses \cite{Guo:06}, the method has evolved rapidly, particularly in the last five years, with notable demonstrations of hybrid achromats and apochromats \cite{Schmid:21}, diffractive lenses \cite{Wang:2020}, flat-optics \cite{doi:10.1021/acs.nanolett.0c04463, Asadollahbaik:2020},millimeter-sized lenses \cite{Ristok20}, catadioptric fiber-optic sensor heads \cite{Lux:24}, and photonic waveguides \cite{Nesic:22} and bundles \cite{Panusa:22}. The possibility of creating sub-mm size freeform optics and/or optical systems is proving particularly useful for endomicroscopy. For instance, Gissibl et al. demonstrated multi-lens objectives measuring only \SI{120}{\micro\meter} in diameter \cite{Gissibl2016}, which can be placed directly at the tip of a fiber bundle. 3D nano-printed lenses on fiber-tips were featured in OCT endomicroscopes with circumferential \cite{Li2018}, and forward-looking \cite{MagneticPositionSensing} scanning as well. Our group has shown that a \SI{1}{\milli\meter} cubic phase plate printed directly on top of a GRIN lens can be used to generate a biaxially accelerated static Airy light sheets for axial sectioning in biomedical imaging \cite{10.1117/1.APN.2.5.056005}. Enabled by a 3D nano-printed "lens-in-lens" structure, Ji et al. developed an intravascular endomicroscopic probe and demonstrated OCT and fluorescence imaging of a mouse artery \textit{in vivo } \cite{Li:2022}. 

While 3D nano-printing of micro-optics is becoming a staple tool, its potential for dynamic micro-opto-electro-mechanical systems (MOEMS) remains relatively unexplored. For miniaturized imaging systems such as multimodal endomicroscopes and multi-photon miniscopes for neural imaging on freely-moving elements, MOEMS are used for performing essential functions including remote focusing for axial scanning \cite{Zong2021, Zong2022}. In addition to the complete design freedom it offers, the inherent mechanical alignment between optical and mechanical components in 3D nano-printing can be a particular advantage over conventional clean-room-based manufacturing processes, where optical components are typically fabricated separately, and integrated with the actuators through precision micro-assembly \cite{Liu2014}, polymer curing \cite{Wu2006, Chen2008} or thermal reflow \cite{Yoo2012}. These processes are not only complex, but also prone to alignment errors that accumulate as the micro-optical system becomes more complex. Furthermore, their reliance on time consuming and costly clean-room processes limits the number of practical design iterations. The first monolithically fabricated MOEMS with continuously translating microlens by up to \SI{88.9}{\micro\meter} was demonstrated by Rothermel et al. \cite{RothermelJournal,10.1117/12.2594213}. This device featured a microlens with a diameter of less than \SI{100}{\micro\meter} at the distal end of a spring, which was actuated electromagnetically through the combination of a microchannel capillary-filled with an epoxy resin containing NdFeB microparticles and an external coil. Due to the omni-directional compliance of the spring, however, the axial lens translation was accompanied by strong parasitic off-axis motion. Our group recently demonstrated a bistable microlens actuator that can switch between two stable predefined states for reconfigurable micro-optical systems \cite{Aybuke}. The device featured a three-dimensional intertwined spiral springs that facilitated the confinement of the lens motion in the axial direction, and was switched by electromagnetically actuating a soft-magnet deposited directly on it through current pulses, and did not consume any DC power in rest states.

In this work, we present a 3D nano-printed lens actuator with large optical aperture, ideal for remote focusing in miniaturized imaging systems. The device uses ortho-planar linear motion springs with a monolithically integrated lens and a self-aligned micro-magnet, and is actuated by two external micro-coils with opposite current directions to minimize torsional forces and facilitate uni-axial motion. It uses a sintered NeFEB micro-magnet instead of a nanoparticles dispersed in polymers to maximize mechanical force, and is optimized for low power consumption with a resonant frequency of more than \SI{300}{\hertz} for mechanical robustness. The \SI{1.4}{\milli\meter} diameter lens profile is optimized through several iterations to minimize both surface roughness and shape errors. Despite the large lens, the entire device, including the coils, has a footprint of \SI{5.74}{\milli\meter} in diameter. Following the manufacturing and a detailed analysis of the optical performance, we conduct a thorough characterization of the mechanical performance, including long-term measurements. Particular emphasis is placed on the effects of coil temperature and the viscoelastic properties of the polymer on actuation. Finally, we compare the performance of the 3D nano-printed actuator with that of actuators fabricated using conventional MEMS processes.

\section{Design}
\begin{figure}[ht]
	\begin{center}
		\begin{tabular}{c}
			\includegraphics[width=14cm]{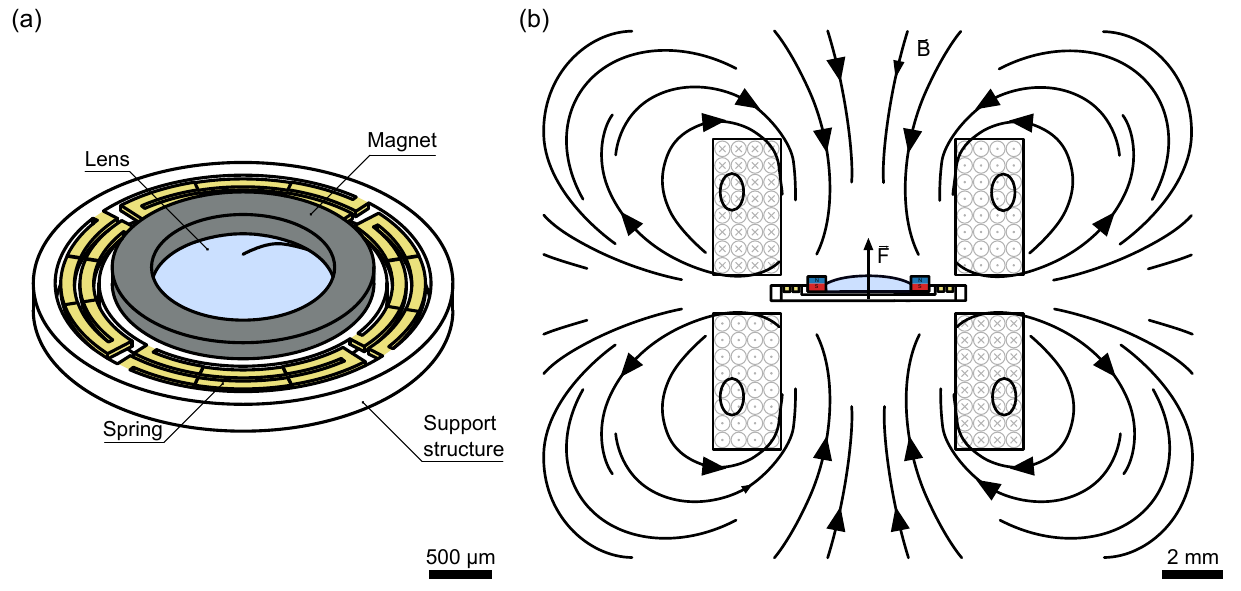}
		\end{tabular}
	\end{center}
	
	\caption[Working principle of the device] {
		\label{fig:functiondrawing_B}
		Working principle of the device. (a) The scanner consists of a magnet, four springs, and a lens. (b) A pair of coils in an arrangement similar to an anti-Helmholtz coil creates a magnetic field gradient at the magnet location. A change in the current through the coils changes this gradient, which in turn changes the force on the actuator. This change in force ultimately causes a displacement of the lens.} 
\end{figure}

The proposed actuator consists of a spring-supported ring-shaped micro-magnet with an integrated lens, positioned between two coils, as depicted in Fig.~\ref{fig:functiondrawing_B}. The electromagnetic gradient generated by the coils exerts a force on the magnet, resulting in a displacement of the magnet-lens assembly along the optical axis.

\subsection{Mechanical design}

The mechanical design is based on ortho-planar linear-motion springs. Four springs connect the lens to the base. Each of the four springs consists of one serpentine spring. Following the nomenclature outlined by Parise et al., this design is referred to as Quad 1-1SC \cite{PARISE20011281}. The planar nature of the springs enables the use of different manufacturing methods, while the use of curved springs results in a very compact design. In addition, the raising and lowering of the lens relative to the base is free of rotational movement \cite{PARISE20011281}, which would be advantageous for potential non-rotationally symmetric lens designs. 

The device uses electromagnetic actuation to attain axial lens translation. If an external magnetic field is applied along the symmetry axis of the actuator, the force $F_{z,m}$ created by the micro-magnet along the same axis is given by

\begin{equation}
	\label{eq:force_by_magnet}
	F_{z, m} = \frac{1}{\mu_0}B_r V \frac{\partial B}{\partial z} = \frac{1}{\mu_0}B_r \frac{m_{m}}{\rho_{m}} \frac{\partial B}{\partial z}\,, 
\end{equation}

with $\mu_0$ being the magnetic field constant, $B_r$ the residual flux density of the micro-magnet, $B$ the applied external field, and $V_m$, $m_m$, and $\rho_m$ the volume, the mass, and the density of the magnet, respectively. This magnetic force is balanced by the force $F_{z, s}$ of the spring given by  

\begin{equation}
	\label{eq:force_by_spring}
	F_{z, s} = k_{\textit{eff}} \Delta z~, 
\end{equation}

where $k_{\textit{eff}}$ is the effective spring constant of the four parallel serpentine springs and $\Delta z$ the displacement of the lens. Equating Eq.~\ref{eq:force_by_magnet} and \ref{eq:force_by_spring} yields the displacement $\Delta z$ as

\begin{equation}
	\label{eq:displacement_simpel}
	\Delta z = \frac{B_r V}{\mu_0 k_{\textit{eff}}}\frac{\partial B}{\partial z}~. 
\end{equation}

As high power dissipation from Joule heating in the electromagnetic coil is one of the main disadvantages of electromagnetic MEMS actuation, this design aims to minimize power consumption for a target lens displacement. As the dissipated power is proportional to the square of the current through the coil and the current is proportional to the required magnetic field, it becomes evident that minimizing the required magnetic field gradient minimizes power dissipation. At the same time, a sufficiently high resonant frequency of the actuator is required to achieve high speed and resilience against environmental disturbances. To include this requirement in the design process, we model the actuator as a simple harmonic oscillator with the resonant frequency $f_0$ given by

\begin{equation}
	\label{eq:simpel_harmonic_oscillator}
	f_0=\frac{1}{2\pi}\sqrt{\frac{k_{\textit{eff}}}{m}}=\frac{1}{2\pi}\sqrt{\frac{k_{\textit{eff}}}{m_{l}+m_{m}}}~, 
\end{equation}

with $m$ being the mass, $m_{l}$ the mass of the lens, and $m_{m}$ the mass of the magnet. Solving Eq.~\ref{eq:simpel_harmonic_oscillator} for $k_{\textit{eff}}$ and substituting it into Eq.~\ref{eq:displacement_simpel}, allows us to define the magnetic field gradient $\tilde{\frac{\partial B}{\partial z}}$  required to achieve certain displacement $\Delta z$ for a design with a defined resonant frequency $f_0$ as

\begin{equation}
	\label{eq:required_magnetic field}
	\tilde{\frac{\partial B}{\partial z}}=\frac{(2\pi f_0)^2(m_{l}+m_{m})\Delta z \mu_0 \rho_m}{B_rm_{m}}~.
\end{equation}

The following two conclusions can be drawn to minimize the required magnetic field gradient.
First, the mass $m_m$ of the magnet must be substantially greater than the mass $m_l$ of the lens. This requirement arises from the fact that the term $\tilde{\frac{\partial B}{\partial z}} \sim \frac{m_l+m_m}{m_m}$ converges to 1 under these conditions.
Second, the ratio of the residual flux density $B_r$ of the magnet to its density $\rho_m$ should be maximized, as $\tilde{\frac{\partial B}{\partial z}} \sim \frac{\rho_m}{B_r}$. Recently, magnets fabricated by filling cavities filled with a compound based on NdFeB microparticles and a low-viscosity 2-component epoxy have been demonstrated \cite{RothermelJournal,Aybuke}. Calikoglu et al. reached a residual flux density $B_r$ of \SI{300}{\milli\tesla} at a density $\rho$ of \SI{5.1}{\gram\per\centi\meter\cubed} \cite{Aybuke}. Compared to the use of compound magnets, the use of conventionally manufactured magnets ($B_r$\,=\,\SI{1.4}{\milli\tesla}, $\rho$\,=\,\SI{7.6}{\gram\per\centi\meter\cubed}) reduces the power consumption by a factor of approximately 10.

The actuator designed in this work has a lens aperture of \SI{1.4}{\milli\meter}. Assuming a lens mass of approximately \SI{0.4}{\milli\gram} using IP-S photoresin, the mass of the magnet is set to \SI{3}{\milli\gram} to fulfill the design condition of $m_m >> m_l$. The micro-magnets, custom machined from pressed and sintered blocks of NdFeB VAC745HR (Audemars Microtec), reach a residual flux density $B_r$ of \SI{1.44}{\tesla} at a density $\rho$ of \SI{7.6}{\gram\per\centi\meter\cubed}. The micro-magnets have an inner diameter of \SI{1.4}{\milli\meter}, an outer diameter of \SI{2}{\milli\meter}, and a height of \SI{250}{\micro\meter}. When determining the geometrical dimensions of the springs and thus getting $k_{\textit{eff}}$, the goal is to achieve a resonant frequency of \SI{300}{Hz} of the first natural mode, which aligns with the intended actuation mode. By choosing \SI{300}{Hz}, we ensure that the first natural mode is more than three times the frequencies commonly found in the ambient environment, which are typically below \SI{100}{\hertz} \cite{ROUNDY20031131}. This separation makes the actuator inherently robust against external movements and vibrations.

The design was optimized using finite element analysis (FEA) with COMSOL Multiphysics\textsuperscript{\tiny\textregistered} 5.6 Structural Mechanics module, assuming a Young's modulus of \SI{5.1}{\giga\pascal}\,\cite{NanoGuide} for the IP-S resin and including geometric non-linearities. Nonlinear and viscoelastic material properties were not taken into account in the simulations. The optimization aimed not only to achieve a resonant frequency of \SI{300}{\hertz}, but also to achieve a linear relationship between force and deflection while ensuring that the stress remains below the tensile strength of IP-S, given by \SI{65}{\mega\pascal} \cite{NanoGuide}.

\begin{figure}[htb]
	\begin{center}
		\begin{tabular}{c}
			\includegraphics[width=14cm]{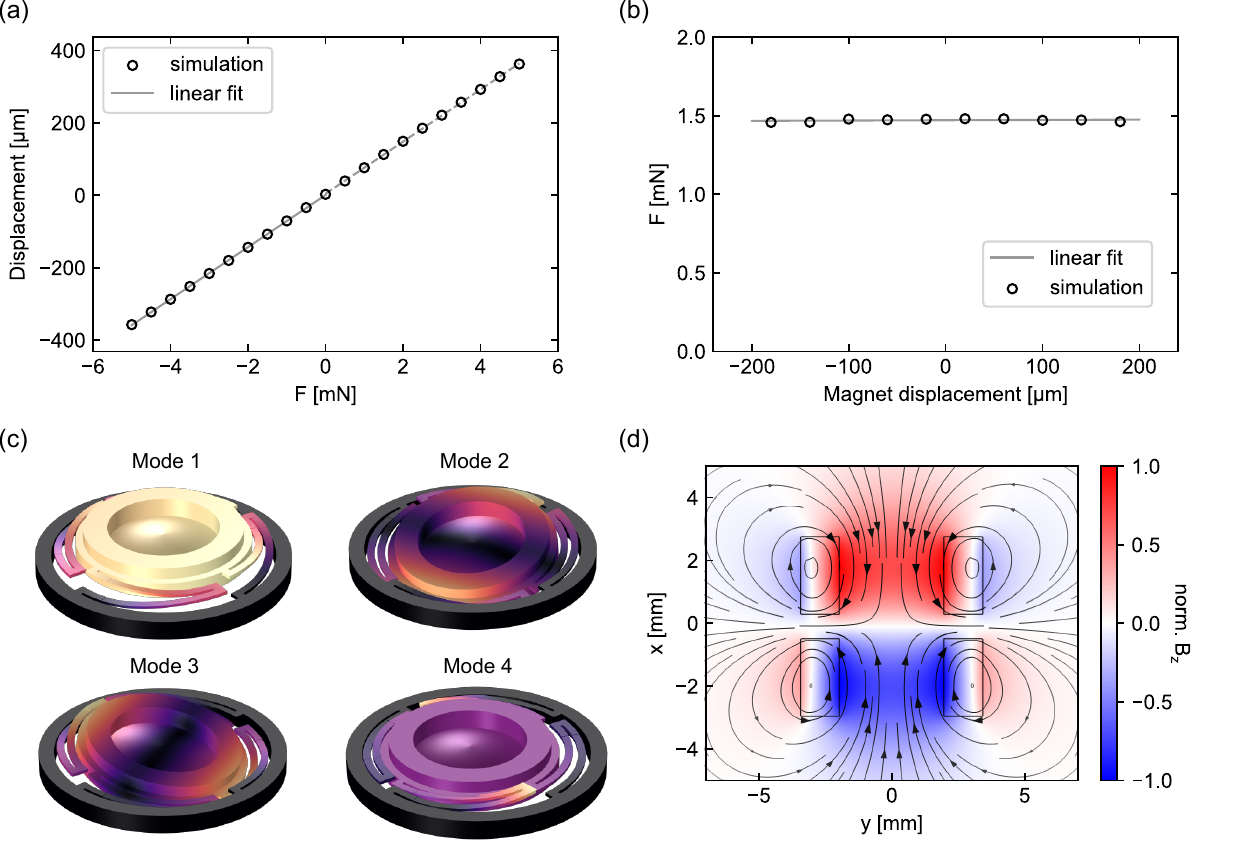}
		\end{tabular}
	\end{center}
	\caption[Mechanical FEA analysis results] 
	{ \label{fig:forcefielddisplacement}
		Mechanical FEA analysis results of the 3D nano-printed MEMS actuator. (a) Simulated displacement of the lens as a function of the applied magnetic force, showing a linear relationship. (b) The simulated force remains constant for all magnet displacements along the optical axis. (c) Resonant modes of the actuator. The first mode is at \SI{303}{\hertz}. (d) Normalized z-component of the simulated magnetic field. The field is approximately zero at the location of the actuator.}
\end{figure} 

The optimized beam thickness is \SI{44}{\micro\meter}, with a reduced thickness of \SI{34}{\micro\meter} at the spring center to reduce stress concentrations in regions prone to maximum strain during deformation. Detailed mechanical dimensions as depicted in Fig.~\ref{fig:techdraw} are given in Tab.~\ref{tab:mech_dimensions}. The simulated displacement $\Delta z$ of the lens as a function of force $F$ is shown in Fig.~\ref{fig:forcefielddisplacement}(a). Until the target displacement range of \textpm~\SI{100}{\micro\meter}, the geometrical nonlinearities remain negligible. The corresponding spring constant $k$ is \SI{13.79}{\newton\per\meter}. The maximum von Mises stress is \SI{29}{\percent} of the yield strength of IP-S. The first 4 natural modes of the actuator depicted in Fig.~\ref{fig:forcefielddisplacement}(c) correspond to resonant frequencies of \SI{303}{\hertz}, \SI{464}{\hertz}, \SI{473}{\hertz}, and \SI{2105}{\hertz}.

\subsection{Coil Design}
A magnet under an external magnetic field does not only experience a force, but also a torque directed to align the internal magnetization vector with the external field. For axial displacement of the lens, only the force along the symmetry axis of the actuator is desired, as any torque tilts the lens with respect to the optical axis, potentially introducing aberrations. As the torque $\vec{\tau}$ is given as

\begin{equation}
	\label{eq:torque}
	\vec{\tau}=V\vec{M}\times \vec{B}~, 
\end{equation}

with $V$ being the volume of the magnet, $\vec{M}$ the magnetization vector, and $\vec{B}$ the magnetic field vector, it becomes evident that the torque becomes zero if either $\vec{B}$ and $\vec{M}$ are parallel or antiparallel to each other, or $\vec{B}$ is equal to zero. As the first condition would require perfect alignment of the magnet to the magnetic field, the second condition was implemented by using two coils with opposite winding directions, similar to an anti-Helmholtz coil. In this configuration, the magnetic fields produced by the two coils cancel each other out at the midpoint, creating a near-zero magnetic field within the actuation region. Consequently, torque is minimized due to the negligible field magnitude, while the force is improved due to the increased magnetic field gradient \cite{Aybuke}.

We optimized the coil pair to minimize power consumption. Since the inner radius of the coils is determined by the size of the actuator and the power consumption is independent of the wire diameter, the only variables that require optimization are the number of axial and radial windings for a given wire diameter. 

The total magnetic field $B(z)$ can be computed by applying the Biot–Savart law to each conductor loop of the coil. It is given as

\begin{equation}
	\label{eq:Biosavart}
	B(z)= \sum\limits_{i=0}^{n} \sum\limits_{j=0}^{m} \frac{\mu_0}{2}\frac{(R_0+r_c+jd_c)^2I}{\big((R_0+r_c+jd_c)^2+(z+r_c+id_c)^2\big)^{3/2}}~, 
\end{equation}

where $R_0$ is the inner radius of the coil, $I$ is the current through the coil, and $z$ is the position along the symmetry axis of the coil relative to the center. $r_c$ and $d_c$ denote the radius and the diameter of the copper wire including the enamel, and $n$ and $m$ denote the number of axial and radial windings, respectively. The coils are made from enameled copper wire with a copper diameter of \SI{114}{\micro\meter}. We determined the fill factor (i.e. the ratio of total conductor area to total coil area in the radial cross-section) of a typical coil to be \SI{70}{\percent} (See Fig.\,\ref{fig:appendix:coilpolish} in the appendix). Therefore, we assume that $d_c$ is \SI{120}{\micro\meter} to match the calculation with the empirically determined fill factor for hexagonal packing. For 20 axial and 10 radial windings, the power consumption reaches a local minimum (see Fig.\,\ref{fig:appendix:coiloptimization} in the appendix). The resistance $R_{coil}$ of this coil pair is 11.5~$\Omega$. The peak current $I_{\textit{peak}}$ required for reaching the required gradient of \SI{3.03}{\tesla\per\meter} for an actuation range of \textpm~\SI{100}{\micro\meter} is \SI{84}{\milli\ampere} at a peak power consumption of \SI{81}{\milli\watt}. The average power when driving the actuator with a triangular waveform is \SI{27}{\milli\watt}. The weight of the coil pair including the coil holder is \SI{505}{\milli\gram}.

The optimized coil design was verified using FEA in the COMSOL Multiphysics \textsuperscript{\tiny\textregistered} AC/DC module. We simulated the force created by the micro-magnet for a current of \SI{84}{\milli\ampere} at different positions along the symmetry axis. As shown in Fig.\,\ref{fig:forcefielddisplacement}(b), the simulated force is constant around \SI{1.48}{\milli\newton} within the actuation range, which is close to the calculated force of \SI{1.38}{\milli\newton} required for a displacement of \textpm\,\SI{100}{\micro\meter}. The simulated force differs from the calculated force since the latter assumes a homogeneous magnetization across the entire magnet volume. In contrast, the simulation accounts for the finite geometrical shape of the magnet as well as spatial variations of the gradient within the coil pair (see Fig.\,\ref{fig:appendix:gradient_simulation} in the appendix).

\subsection{Lens Design}
In principle, any optical component that can be fabricated using two-photon polymerization can be integrated into the actuator. In this work, we opted for a plano-convex lens designed to focus collimated light with a wavelength of \SI{850}{\nano\meter}, and a possible solution to perform remote focusing in a multi-photon miniscope, such as the one discussed in \cite{Zong2022}. We used   ZEMAX\textsuperscript{\textregistered} OpticStudio\textsuperscript{\textregistered} for the design and optimization of the lens, which has a diameter of \SI{1.4}{\milli\meter}, matching the inner diameter of the magnet, and an image space $\textit{NA}$ of 0.2, resulting in a back focal length of \SI{3.42}{\milli\meter}. The lens surface is aspherical with a radius $R$ of \SI{-1.72}{\milli\meter} and a conic constant $\kappa$ of \SI{-2.25}{}. The lens is optimized for IP-S photoresin with a refractive index of \SI{1.5}{} \cite{Gissibl17}.

\section{Fabrication}
\begin{figure}[ht]
	\begin{center}
		\begin{tabular}{c}
			\includegraphics[width=14cm]{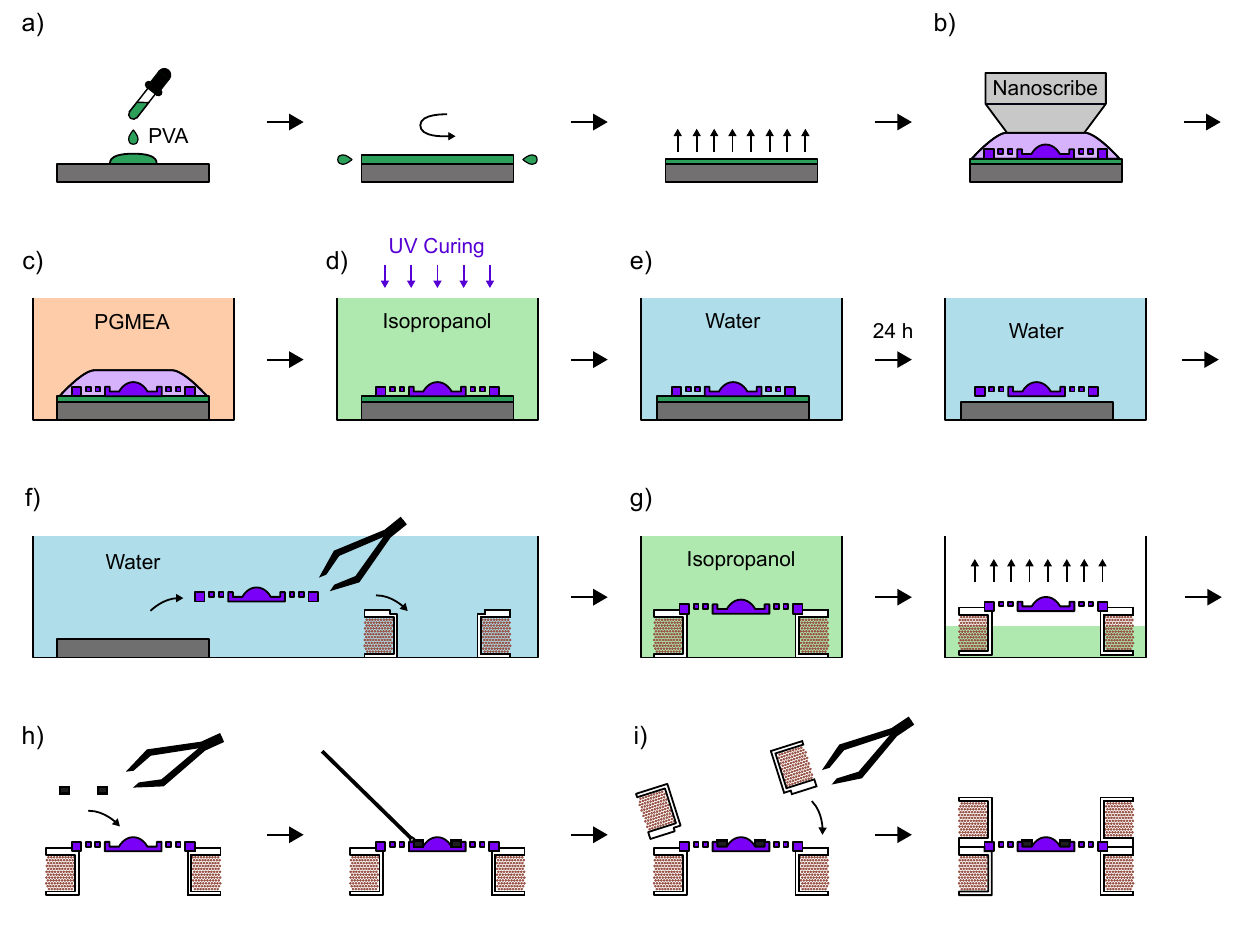}
		\end{tabular}
	\end{center}
	\caption[Fabrication process of the complete actuator]
	{ \label{fig:fabrication_scheme}
		Fabrication process of the complete actuator. A \SI{20}{\nano\meter} layer of PVA is spin-coated onto the substrate (a) prior to 3D nano-printing (b). The structure is developed in PGMEA (c) and rinsed in IPA under UV irradiation (d). PVA is dissolved in water to enable lift-off. (e) After lift-off, the structure is placed in the lower coil (f). Before the evaporation of the solvent, water is exchanged by IPA to lower the capillary forces (g). The micromagnet is placed in the actuator and fixed using UV-curable adhesive (h). The second coil is placed on the actuator and fixed using UV-curable adhesive (i).}
\end{figure}

The complete fabrication process, summarized in Fig.\,\ref{fig:fabrication_scheme}, consists of four sub-processes: substrate preparation, monolithic 3D nano-printing of the actuator including springs and lens, lift-off, and assembly.

\subsection{Substrate preparation}

A \SI{20}{\nano\meter} layer of polyvinylalcohol (PVA, 87-89\,\% hydrolyzed, M\textsubscript{w}\,13,000-23,000, 363081, Sigma-Aldrich) was spin-coated at \SI{1750}{RPM}\, from an aqueous solution (1\,\%\,w/w) onto a silicon substrate using a static dispense (Fig.\,\ref{fig:fabrication_scheme}(a)). 

\subsection{Micro 3D nano-printing and Development}

The actuator is printed in the dip-in configuration using a commercial 3D nano-printer (Nanoscribe GmbH, Photonic Professional Gt+) equipped with a 10x/0.3NA objective using IP-S photoresin (Nanoscribe GmbH) (Fig.\,\ref{fig:fabrication_scheme}(b)). The printing parameters for the mechanical support, springs, and lens are summarized in Tab.~\ref{tab:printing_paramters}. We printed the mechanical parts with a larger slicing distance, since an optical-quality surface is not required for these. The lens was printed in two stages. In the first stage, a coarse print with a large slicing distance allows for a fast printing of the lens volume. In the second stage, a shell with a thickness of \SI{5}{\micro\meter} was printed on top of the already printed lens using a small slicing distance to create an optical-quality surface. This printing strategy shortened the printing time of the lens by a factor of approximately three compared to conventional printing, where the complete volume of the lens would be printed at a small slicing distance. 

To account for shape deviations caused by shrinkage during the printing process, we printed multiple iterations of the lens. After each iteration, the shape was measured and the print file was adapted accordingly using standard Zernike polynomials \cite{Tage2023}. During this process, the lens was printed without springs or mechanical support.

Printing the long, overhanging springs using the standard layer-by-layer approach is not a suitable option, since the layers printed without support are floating, leading to distorted or failed prints\cite{Gross2019}. To overcome this problem, we adapted the printing strategy developed by Marschner et al. \cite{Marschner2023}, where the overhanging structure is divided into several small parallelepipeds that are printed sequentially along the radius of the spring. We used a block length of \SI{20}{\micro\meter} and a shear angle of \SI{15}{\degree}. Individual blocks overlap by \SI{4}{\micro\meter}. We developed a custom Python script for General Writing Language (GWL) programming. The GWL file defines the path the laser focus will trace within the resin. To avoid spring deformation due to capillary forces during developer evaporation, which might push the springs beyond their yield strength and potentially cause plastic deformation or destruction \cite{Namatsu1999}, we constrained the movement of the springs by connecting them to the passive mechanical structure of the actuator using safety pins. Those pins have a height of \SI{20}{\micro\meter} and a width of \SI{5}{\micro\meter}.   

Following printing, we developed the actuator in propylene glycol monomethyl ether acetate (PGMEA) for 2 hours (Fig.\,\ref{fig:fabrication_scheme}(c)), before rinsing with isopropyl alcohol (IPA) for 30 minutes to remove the PGMEA. To increase the cross-linking and reduce aging effects, we flood illuminated the structure with \SI{365}{\nano\meter} light (Dr.\,Hönle\,AG, LED Pen 2.0), while still being submerged in IPA \cite{Schmid:19,Purtov2018} (Fig.\,\ref{fig:fabrication_scheme}(d)). 

\begin{table}[htb]
	\caption[Parameters used for printing]{Parameters used for printing the actuator from IP-S photoresin using the Nanoscribe Photonic Professional Gt+ equipped with a 10x/0.3NA objective.} 
	\label{tab:printing_paramters}
	\vspace{10pt}
	\centering
	\def\arraystretch{1.5}
	\setlength{\tabcolsep}{4pt}
	\begin{tabular}{llcccc}
		\hline
		&  & support & springs & lens & lens shell \\ \cline{1-1} \cline{3-6}
		slicing distance [\SI{}{\micro\meter}]  &  & 1.5 & 2 & 1.5 & 0.2  
		\\ \cline{1-1} \cline{3-6} 
		hatching distance [\SI{}{\micro\meter}] &  & 0.5  & 1.0 & 0.5 & 0.5            \\ \cline{1-1} \cline{3-6} 
		power [\SI{}{\milli\watt}]             &  & 54.5 & 50 & 54.5  &54.5          \\ \cline{1-1} \cline{3-6} 
		scan speed [\SI{}{\milli\meter\per\second}]        &  & 100 &100   & 100 & 35         \\ \cline{1-1} \cline{3-6} 
		mode              &  & \multicolumn{4}{c}{solid}        \\ \hline
	\end{tabular}
\end{table}

\subsection{Lift-off}

The substrate is immersed in water to dissolve the PVA, resulting in the lift-off of the 3D nano-printed structure from the substrate (Fig.\,\ref{fig:fabrication_scheme}(e)). While still immersed in water, the actuator is placed onto the lower electromagnetic coil (Fig.\,\ref{fig:fabrication_scheme}(f)). The coil holders are fabricated out of fused-silica using selective laser-induced etching (SLE) using a commercial laser microscanner (LightFab GmbH). We used enameled copper wire (1570225, TRU Components, Conrad Electronics SE, Germany) and a custom winding machine to create the coil using orthocyclic winding. Water is exchanged by IPA to reduce capillary forces during the evaporation of the solvent (Fig.\,\ref{fig:fabrication_scheme}(g)). 

\subsection{Assembly}

\begin{figure}[ht]
	\begin{center}
		\begin{tabular}{c}
			\includegraphics[width=14cm]{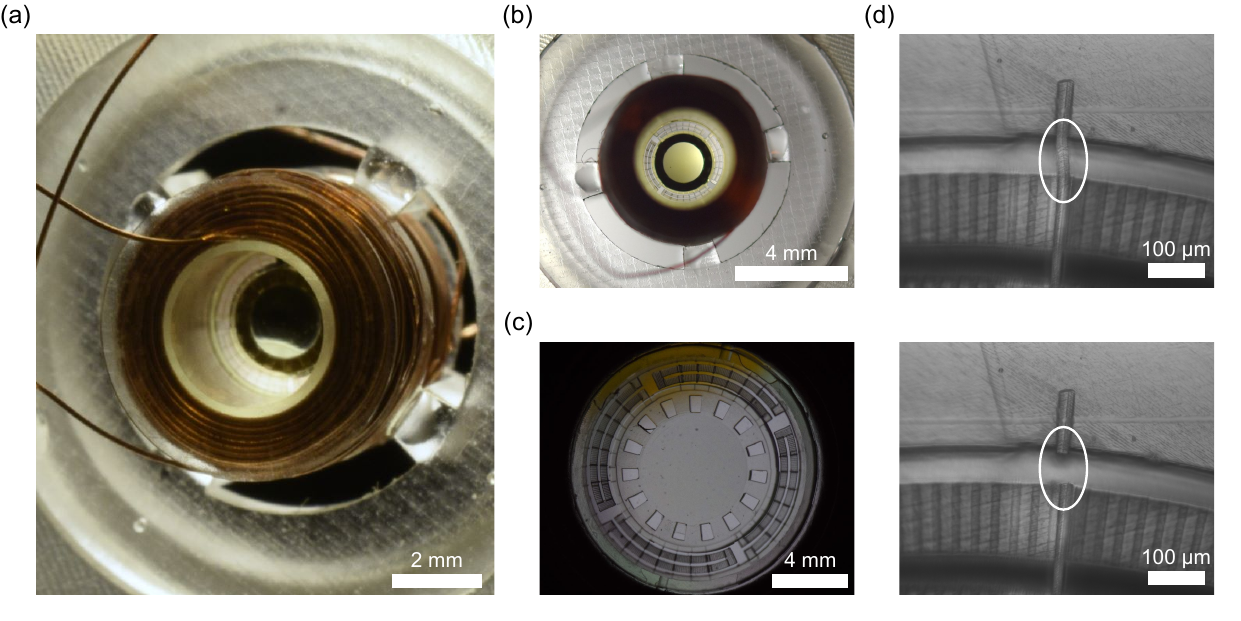}
		\end{tabular}
	\end{center}
	\caption[Photographs of the actuator] 
	{ \label{fig:images_of_actuator}
		Photographs of the actuator. (a,b) Complete actuator with coil pair. (c) Actuator before magnet integration. (d) Mechanical spring before and after the ablation of the safety pins.} 
\end{figure}

Following the lift-off, we removed safety pins constraining the springs by femtosecond laser multi-photon ablation \cite{Xiong2012} using the same commercial laser microscanner (LightFab GmbH) equipped with a 
20x/0.45NA objective (DIC N1 OFN22 , Nikon Inc.) at a power of \SI{100}{\milli\watt} and a scan speed of \SI{80}{\micro\meter\per\second}. Figure~\ref{fig:images_of_actuator} provides a close-up view of the springs before and after beam removal. The micro-magnet is placed in the actuator and secured using UV-curable adhesive (Panacol Vitralit\textsuperscript{\textregistered} UC 1618) (Fig.\,\ref{fig:fabrication_scheme}(h)). Subsequently, the second coil is placed on the actuator and secured using the same glue (Fig.\,\ref{fig:fabrication_scheme}(i)). Passive alignment structures ensure alignment (\textpm\,\SI{20}{\micro\meter}) of the two coils relative to each other.  

\newpage

\section{Mechanical characterization}

For mechanical characterization, the coils were driven by a custom printed circuit board, implementing a bidirectional voltage-controlled current source (Howland current source). The drive signal was generated using an arbitrary function generator (Tektronix GmbH, AFG1022) and the displacement was measured using a single-point laser Doppler vibrometer (Polytec GmbH, VGo-200).

\subsection{Frequency response} \label{Results_Fequency_Response}

\begin{figure}[ht]
	\begin{center}
		\begin{tabular}{c}
			\includegraphics[width=14cm]{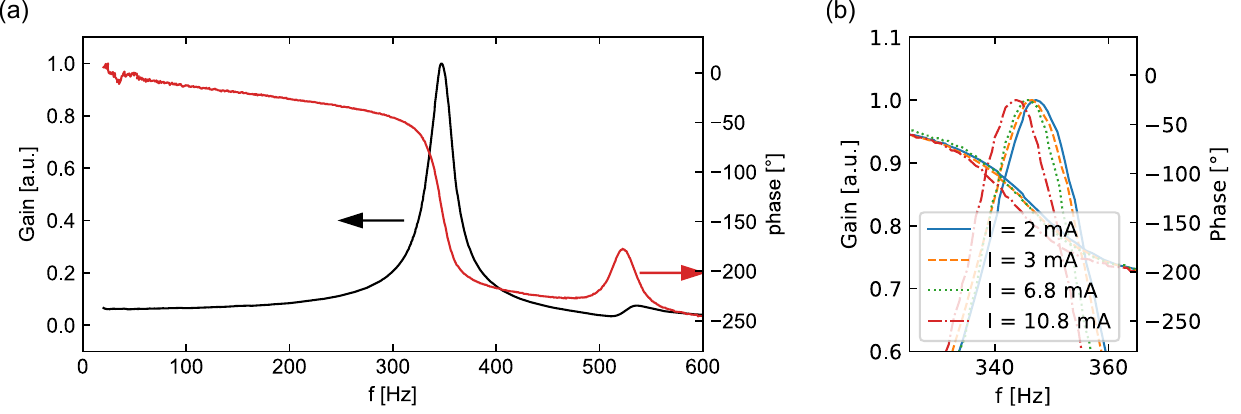}
		\end{tabular}
	\end{center}
	\caption[Frequency response measurement] 
	{ \label{fig:ac_response}
		Frequency response of the actuator. (a) Complete frequency response for a peak current of \SI{2}{\milli\ampere}. The resonant frequency of the first mode is \SI{347.3}{\hertz} with a quality factor \SI{17}{}. (b) The resonant peak of the fundamental mode for different drive currents. A shift from \SI{347.3}{\hertz} to \SI{343.8}{\hertz} is observed for an increase of the drive current from \SI{2}{\milli\ampere} to \SI{10.8}{\milli\ampere}.
	}
\end{figure}

To evaluate the frequency response of the actuator, the input signal was swept from \SI{10}{\hertz} to \SI{700}{\hertz} over a period of \SI{10}{\second}. This measurement was repeated for different drive currents. Figure\,\ref{fig:ac_response} shows the complete frequency response for a drive current of \SI{2}{\milli\ampere}, and the first resonant peak for four different drive currents. The resonant frequency of the first mode is \SI{347.3}{\hertz}, which is 12\,\% higher than the design value. The device gain (i.e. displacement per unit drive current) is \SI{0.91}{\micro\meter\per\milli\ampere}, which is 24\% smaller than the simulated result of \SI{1.19}{\micro\meter\per\milli\ampere}. 

Both these results point to the spring constant of the device being larger than the design specifications, which is most likely due to the strong dependence of the Young's modulus on the curing parameters \cite{youngsmodulus_two_photon, mi8040101}. The quality factor is \SI{17}{}. For increasing drive currents, the resonant peak shifts towards lower frequencies, e.g. from \SI{347.3}{\hertz} for \SI{2}{\milli\ampere} to \SI{343.8}{\hertz} at \SI{10.8}{\milli\ampere}, indicating spring softening. At \SI{10.8}{\milli\ampere}, the measured peak-to-peak displacement was \SI{172}{\micro\meter}.

\subsection{Quasi-static actuation}
\label{section_quasistatic_actuation}

\begin{figure}[ht]
	\begin{center}
		\begin{tabular}{c}
			\includegraphics[width=14cm]{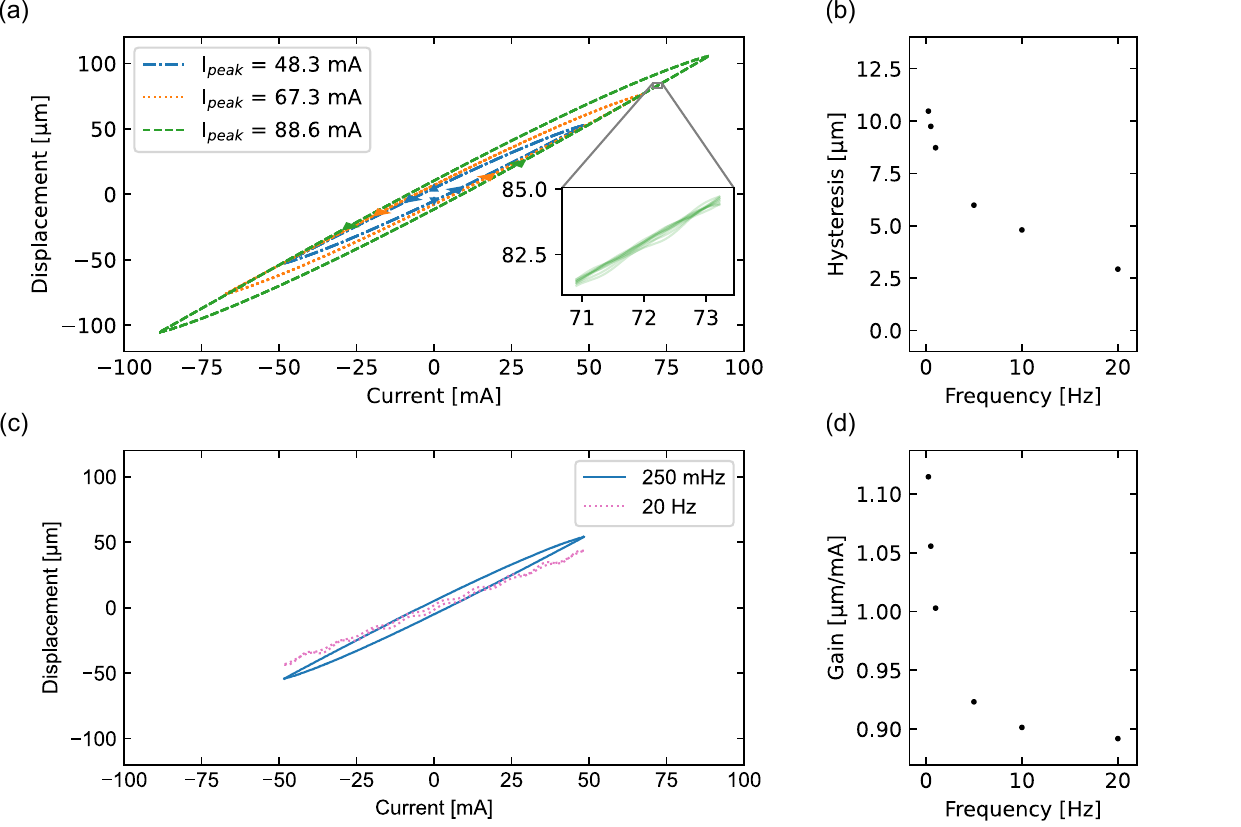}
		\end{tabular}
	\end{center}
	\caption[Quasi-static actuation measurement]
	{ \label{fig:dc_response}
		Quasi-static actuation characteristics of the MEMS scanner under different conditions. (a) Quasi-static displacement of the MEMS actuator as a function of applied current for three peak current levels at \SI{1}{\hertz}. Significant hysteresis is observed. (b) The hysteresis amplitude decreases with increasing frequency, measured in the range from \SI{250}{\milli\hertz} to \SI{20}{\hertz}. (c) Displacement versus current profiles for two selected drive frequencies show reduced hysteresis at higher frequencies. (d) The gain decreases with increasing frequency.} 
\end{figure}

To evaluate the characteristics of quasi-static displacement, the actuator was driven using a triangular signal at \SI{1}{\hertz} with three different peak currents. At a peak current of \SI{88.6}{\milli\ampere}, a peak-to-peak displacement of \SI{209.2}{\micro\meter} was observed. The device response to the triangular wave has significant hysteresis, as shown in Fig.\,\ref{fig:dc_response}a for three different drive currents. To quantify the hysteresis and investigate its origin, the drive frequency was varied between \SI{250}{\milli\hertz} and \SI{20}{\hertz} for a peak current of \SI{48.3}{\milli\ampere}. Two sample datasets are presented in Fig.\ref{fig:dc_response}c. Figure\,\ref{fig:dc_response}b illustrates the hysteresis as a function of drive frequency, with a decrease towards higher frequencies. For instance, at \SI{250}{\milli\hertz}, the hysteresis is \SI{10.5}{\micro\meter}, while at \SI{20}{\hertz}, it decreases to \SI{2.9}{\micro\meter}. With \SI{1.11}{\micro\meter\per\micro\ampere} at \SI{250}{\milli\hertz} and \SI{0.89}{\micro\meter\per\milli\ampere} at \SI{20}{\hertz}, the gain follows a comparable trend, as shown in Fig.\,\ref{fig:dc_response}d.

The hysteresis behavior and its evolution with excitation frequency can be explained by the viscoelastic material properties of the IP-S photoresin. At low drive frequencies, such as \SI{250}{\milli\hertz}, the period of the drive signal is comparable to the viscoelastic time constant of the material, allowing both elastic and viscoelastic effects to contribute to the observed displacement. At higher frequencies, such as \SI{20}{\hertz}, the period falls well below the viscoelastic time constant \cite{RothermelJournal}, resulting in reduced viscoelastic contributions and thus reduced hysteresis and gain, which approaches the value derived from the frequency response (\SI{0.89}{\micro\meter\per\milli\ampere} vs. \SI{0.91}{\micro\meter\per\milli\ampere}).

\subsection{Step response}
\begin{figure}[ht]
	\begin{center}
		\begin{tabular}{c}
			\includegraphics[width=14cm]{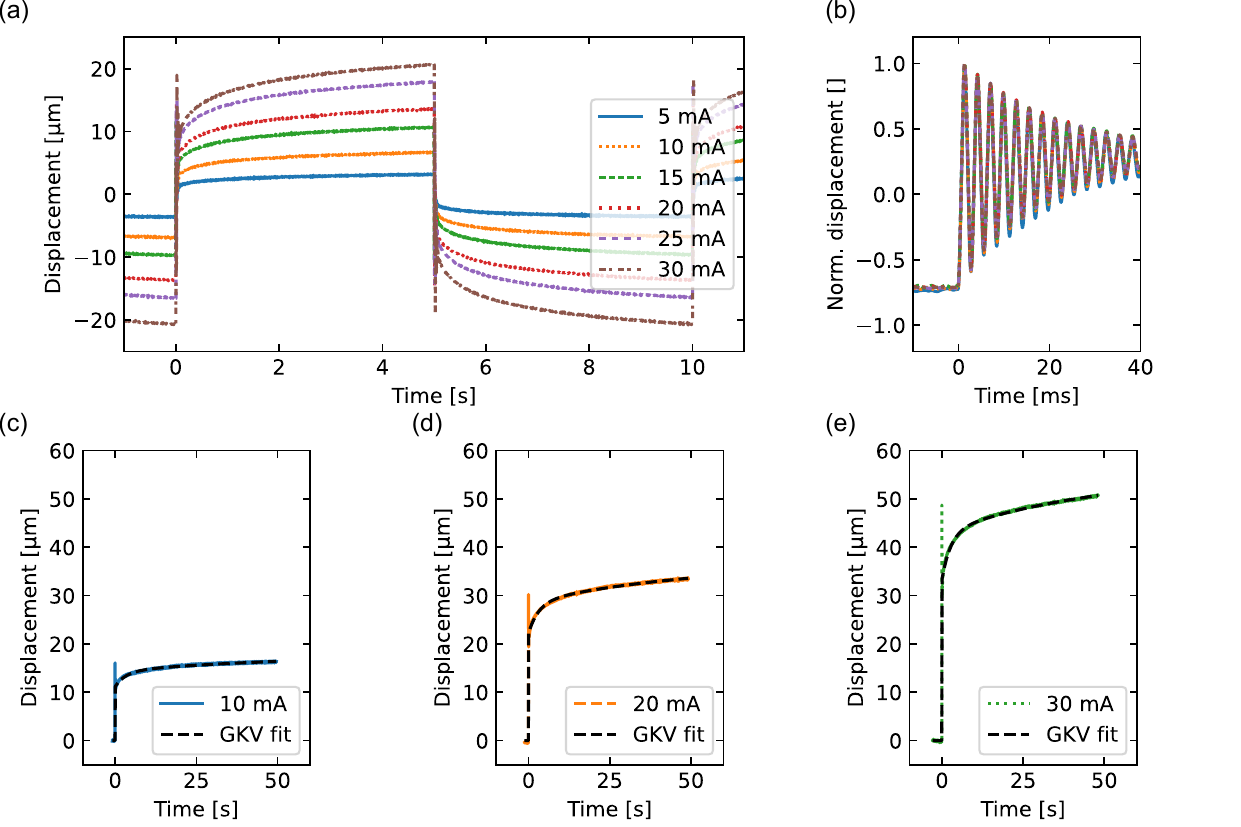}
		\end{tabular}
	\end{center}
	\caption[Step response measurement]
	{ \label{fig:stepsresponse}
		Step response of the MEMS actuator under different driving currents. (a) The displacement of the actuator over time for different current levels. Viscoelastic creep is observable after the transient response. (b) Initial ringing occurs following the step input and stabilizes over time. The rise time $\tau_r$ is \SI{0.51}{\milli\second}\,\textpm\,\SI{0.02}{\milli\second}. (c-e) Step response for three different currents. The data are fitted with a general Kelvin-Voigt viscoelasticity model with two elements \cite{Serra2019}.
	}
\end{figure}

The response time of the device was quantified by measuring the step response at different drive amplitudes. Figure\,\ref{fig:stepsresponse}a depicts the device response to a rectangular signal with bidirectional actuation for different currents. The average rise time $\tau_r$ from 10\,\% to 90\,\% is \SI{0.51}{\milli\second}\,\textpm\,\SI{0.02}{\milli\second}.  Following the initial step, a brief period of ringing is observed as shown in more detail in Fig.\,\ref{fig:stepsresponse}b, since the device works in the under-damped regime. The viscoelasticity is manifested as the creep that follows the ringing of the actuator. 

\begin{figure}[htb]
	\begin{center}
		\begin{tabular}{c}
			\includegraphics[width=7cm]{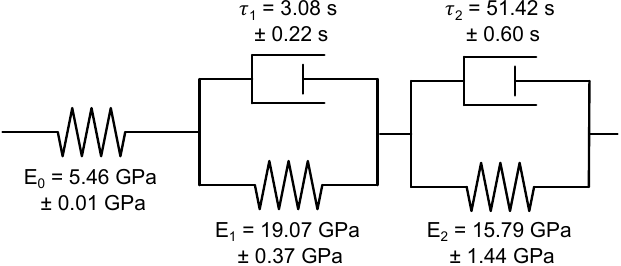}
		\end{tabular}
	\end{center}
	\caption[GKV model]
	{ \label{fig:GKVmodel}
		A general Kelvin-Voigt viscoelasticity model of second order \cite{Serra2019} was used to characterize the viscoelastic material properties. The model was fitted to the step response of the device for three different currents. Errors correspond to the standard deviation of the mean.
	}
\end{figure}

To quantify the viscoelastic material properties, we fitted a general Kelvin-Voigt model (GKV) of second order \cite{Serra2019} to the step responses measured at three different current values, shown in Fig.\,\ref{fig:stepsresponse}(c-d). In the GKV model, which is represented in Fig.\,\ref{fig:GKVmodel}, the displacement $s$ as a function of time $t$ can be described as

\begin{equation}
	\label{eq:compliance_function}
	s(t) = F\left( \frac{C}{E_0 I}+\frac{C}{E_1 I}(1-e^{-\frac{t}{\tau_1}})+\frac{C}{E_2 I}(1-e^{-\frac{t}{\tau_2}})\right), 
\end{equation}

with $F$ being the force, $C$ the geometry dependent constant determined in the simulation, $E$ the Young's modulus and $\tau$ the time constant. The results are summarized in Fig.\,\ref{fig:GKVmodel}. While the time constants are consistent with values reported in the literature, the Young's moduli are higher. These deviations are expected, as the material properties depend on the printing parameters\cite{youngsmodulus_two_photon, mi8040101}, as discussed in Sec.\,\ref{chap:discussion}.

\subsection{Thermal effects}
\begin{figure}[ht]
	\begin{center}
		\begin{tabular}{c}
			\includegraphics[width=14cm]{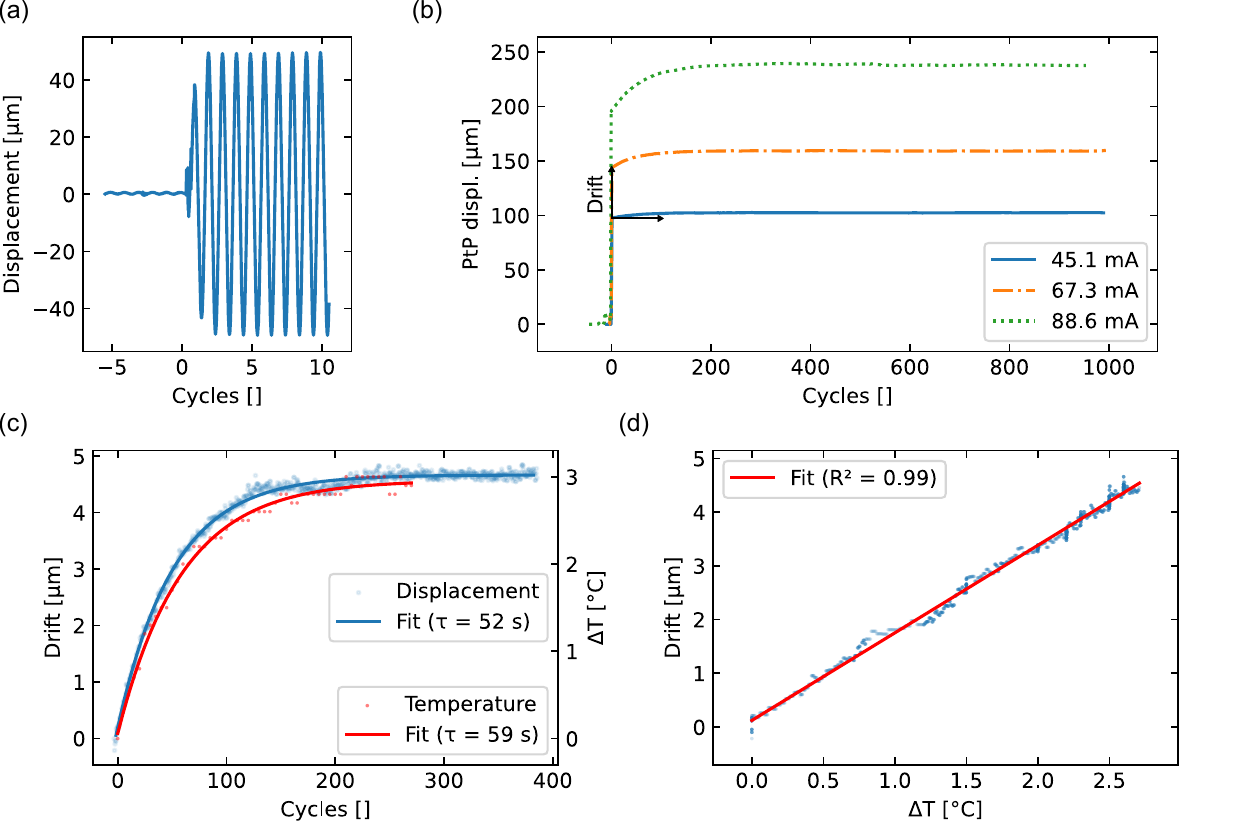}
		\end{tabular}
	\end{center}
	\caption[Measurement of the long-term stability] 
	{ 
		\label{fig:longterm_quasistatic}
		Measurement of the long-term stability. (a) Zoom in to the displacement over time. (b) Peak-to-peak displacement of the actuator over 1000 cycles for three different drive currents. (c) Evaluation of displacement drift and coil temperature over time for a current of \SI{45.1}{\milli\ampere}. (d) Linear correlation between displacement drift and coil temperature} 
\end{figure}

Another important effect that strongly influences the Young's modulus of polymers is temperature. Rohbeck et al. demonstrated the temperature dependence of the Young's modulus for IP-Dip, a photoresin similar to IP-S \cite{ROHBECK2020108977}. This decrease in Young's modulus leads to a reduced spring constant and thus an increased gain. Since electromagnetic actuation is relatively power-hungry compared to methods like electrostatic or piezoelectric actuation, operating the device over extended periods would potentially lead to heating, and thus changing device behavior. To evaluate the long-term stability of the device, we applied a \SI{1}{\hertz} sinusoidal drive signal at different peak currents for approximately 1000 cycles each. Figure\,\ref{fig:longterm_quasistatic}a depicts the displacement as a function of time for the first 10 cycles, while Fig.\,\ref{fig:longterm_quasistatic}b shows the measured peak-to-peak displacement over the full number of cycles. The results indicate that the displacement remains stable over time. However, it is apparent that the shift increases during the first few hundred cycles before stabilizing at a constant value.

Figure\,\ref{fig:longterm_quasistatic}c illustrates the thermal drift, defined as the additional peak-to-peak displacement relative to the initial value, along with the coil temperature over time, which was measured using thermal imaging (ETS320, Teledyne FLIR LLC). The time constant for the drift of \SI{52}{\second} closely matches the time constant of the coil temperature of \SI{59}{\second}. There is a strong linear correlation between drift and coil temperature (R\textsuperscript{2}\,=\,0.99), as depicted in Fig.\ref{fig:longterm_quasistatic}d. Similar analyses for increased drive currents show comparable results (see Fig.\,\ref{fig:appendix:longtermtemperature} in the appendix).

\section{Optical characterization}
To examine the characteristics of the monolithically integrated lens, we first measured its shape, followed by an optical characterization.

\subsection{Lens Shape}
The lens shape was measured using a 3D optical profiler (ZYGO NewView™ 9000). We measured the root-mean-square (RMS) roughness within a square region of 25\,x\,\SI{25}{\micro\meter\squared}, applying a Gaussian high-pass filter with a cutoff period of \SI{25}{\micro\meter} (\emph{EN ISO 25178:2012}). Standard Zernike polynomials were fitted to both the measured and ideal surfaces to quantify the shape deviation \cite{Tage2023}.

\begin{figure}[ht]
	\begin{center}
		\begin{tabular}{c}
			\includegraphics[width=14cm]{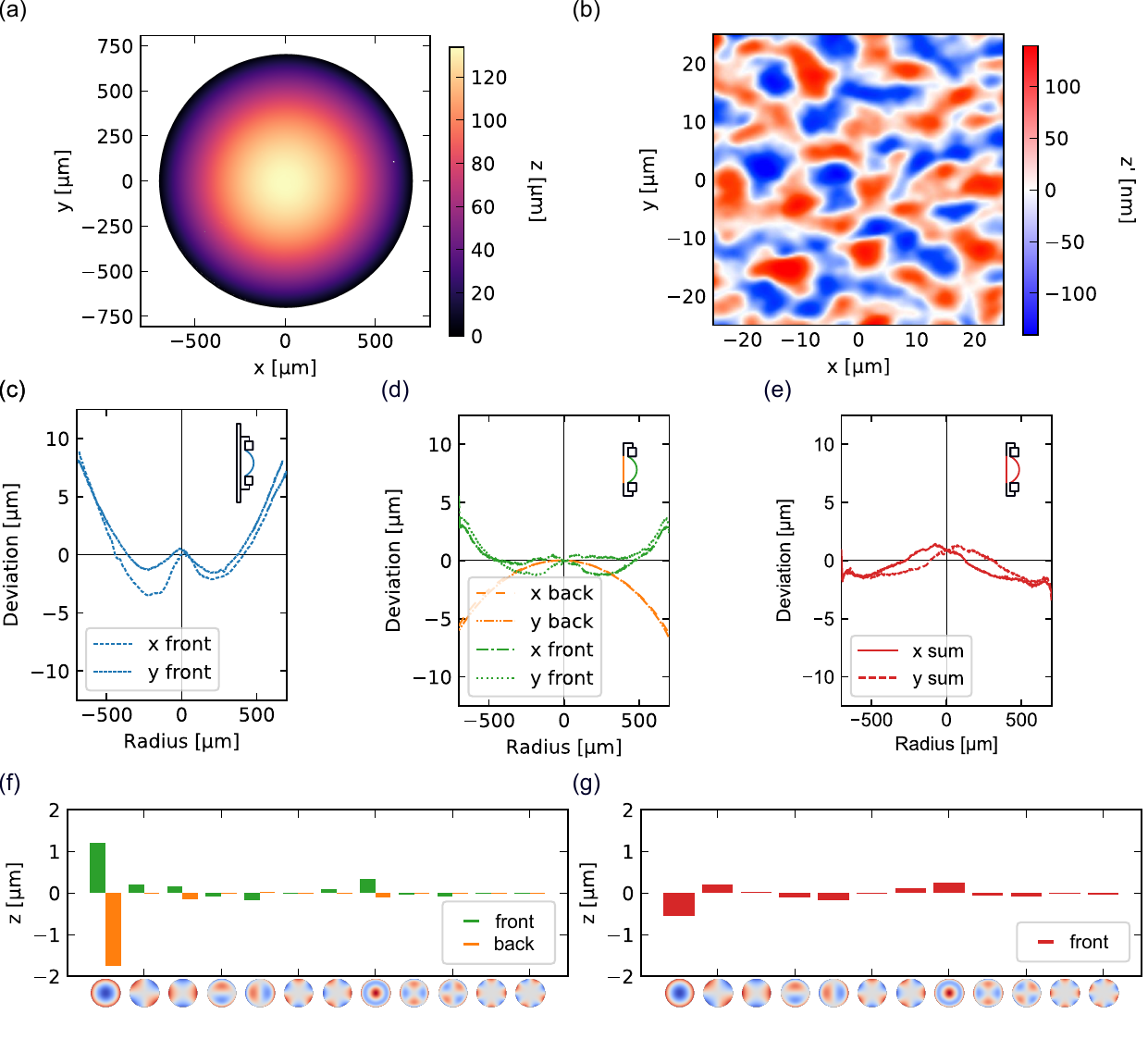}
		\end{tabular}
	\end{center}
	\caption[Results of the shape characterization]
	{ \label{fig:shape_data}
		Results of the shape characterization. (a) A complete profile of the printed lens. (b) Gaussian high-pass filtered section of the lens. The RMS roughness is \SI{37}{\nano\meter}. (c) Shape deviation of the uncompensated, non-released lens. (d) Shape deviation of the back and the front surface of the lens after lift-off. (e) Total shape deviation (thickness deviation) of the lens. (f) Zernike polynomials of the front and the back surfaces. (g) Combined Zernike polynomials}
\end{figure}

The results of the shape analysis, presented in Fig.\,\ref{fig:shape_data}, include the complete measured surface with an RMS roughness of \SI{37}{\nano\meter}, as well as cross-sections in the x- and y-directions. The deformation of the lens upon release from the substrate is attributed to residual stress caused by polymer shrinkage during the printing process. Consequently, the deformation of the first (flat) surface is transferred to the second (aspherical) surface. Since the overall thickness profile determines the optical characteristics of the lens, the deviations in the first and second surfaces are combined. Comparing the combined shape deviation to that of the uncompensated lens reveals a significant improvement in shape accuracy. For instance, defocus decreased from \SI{3.30}{\micro\meter} to \SI{-0.55}{\micro\meter}, and spherical aberrations were reduced from \SI{0.47}{\micro\meter} to \SI{0.25}{\micro\meter}. The total RMS shape deviation decreased by 80\% from \SI{4.65}{\micro\meter} to \SI{0.94}{\micro\meter}.

\subsection{Optical Performance}\label{chapter_optical_performance}
To evaluate the optical performance, the actuator was illuminated with a collimated \SI{850}{\nano\meter} pigtailed laser (LDM-850-V-0.2, Bitline System Pty Ltd). The focal plane was imaged using a 0.55\,NA objective lens (Plan Apo 50x, 378-805-3, Mitutoyo AC) in combination with an infinity-corrected tube lens (TTL200-A, Thorlabs, Inc.), and captured with a CCD camera (UI-1240SE-NIR, IDS GmbH).

\begin{figure}[ht]
	\begin{center}
		\begin{tabular}{c}
			\includegraphics[width=14cm]{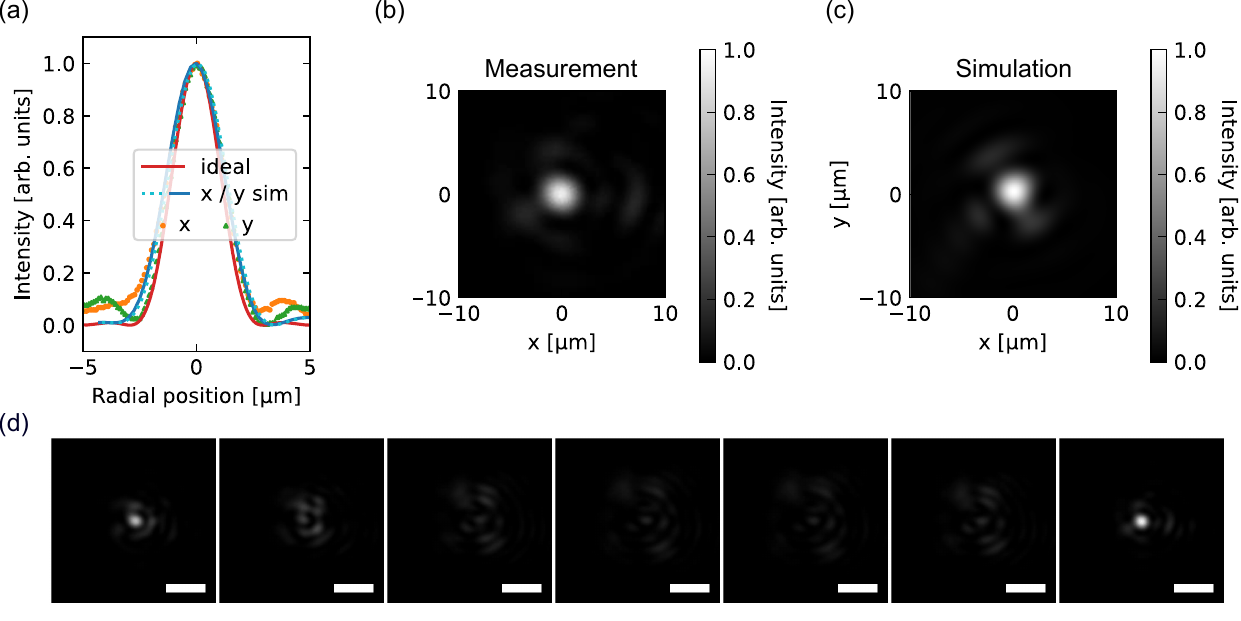}
		\end{tabular}
	\end{center}
	\caption[Results of the optical characterization]
	{ \label{fig:optical_characterization} 
		Results of the optical characterization. (a) Profile of the measured focal spot, the simulated spot based on the measured shapes, and the simulated spot based on the ideal shape. (b) Focal spot of the 3D nano-printed actuator. (c) Simulation of the focal spot based on the measured shape. (d) Still images of the focal plane while the actuator translates the lens along the optical axis. Scale bar: \SI{5}{\micro\meter} (Video\,1, AVI, 0.7\,MB).} 
\end{figure}

The plane of best focus is located at \SI{3.29}{\milli\meter} which is close to the designed value of \SI{3.42}{\milli\meter}. Figure\,\ref{fig:optical_characterization} shows cross sections of the focal spot, as well as focus images of the measurement and the simulation based on the measured lens shape. The spot size (FWHM) is \SI{2.65}{\micro\meter} in the x-direction and \SI{2.54}{\micro\meter} in y-direction. The expected spot size for the designed lens is \SI{2.40}{\micro\meter}. The deviation between the measured and ideal spot size is associated with the shape deviation of the lens, which is in essence a combination of astigmatism, coma, and spherical aberrations. The mismatch between the ideal and measured back focal length is related to the deviation of the defocus.

\section{Discussion}\label{chap:discussion}

The fabricated and characterized actuator achieves a displacement of \textpm\,\SI{100}{\micro\meter}, aligning well with the design specifications. However, both the resonant frequency and gain deviate from the design, with an increase in the resonant frequency and a decrease in gain. This discrepancy is likely attributed to a higher-than-expected spring stiffness, which may arise from geometric mismatches in the spring profile or deviations in the Young’s modulus from the assumed value during design and simulation. Notably, the Young’s modulus of two-photon polymerized materials depends on printing parameters such as laser power, writing speed, hatching distance, and slicing distance \cite{youngsmodulus_two_photon, mi8040101}. A significant portion of the springs was manufactured with overlapping volumes for stitching, resulting in high cross-linking levels. This increased cross-linking correlates with an elevated Young’s modulus \cite{youngsmodulus_two_photon, Pagliano2022}. For instance, Pagliano et al. observed a need to adjust the Young’s modulus from \SI{5.1}{\giga\pascal} to \SI{6.5}{\giga\pascal} to align simulation results with measured resonant frequencies \cite{Pagliano2022}. Nano-indentation tests on the fabricated springs could provide valuable insights into this aspect. Regarding spring thickness, single-point measurements revealed minor deviations (\SI{44.5}{\micro\meter} vs. \SI{44.0}{\micro\meter} and \SI{33.7}{\micro\meter} vs. \SI{34.0}{\micro\meter}). Incorporating these deviations into simulations increased the resonant frequency from \SI{303}{\hertz} to \SI{309}{\hertz}. However, these measurements do not account for potential variations in thickness along the spring length. 

Characterization of the device’s quasi-static actuation and step response revealed hysteresis and mechanical creep, likely due to the viscoelastic properties of the material. Advances in material optimization could mitigate these issues, reducing hysteresis and improving mechanical performance. The optical performance of the printed lens closely matches the design, with residual shape deviations causing slight discrepancies in spot size and focal length. Further iterations of shape compensation could refine the optical accuracy \cite{Ristok20}, bringing measurements and design into closer alignment.

When compared to conventionally manufactured lens actuators, which achieve scanning ranges of up to \SI{480}{\micro\meter} with electrostatic actuation \cite{Li2019}, \SI{418}{\micro\meter} with electromagnetic actuation \cite{Ou2023}, \SI{400}{\micro\meter} with thermal actuation \cite{Liu2014}, and \SI{430}{\micro\meter} with piezoelectric actuation \cite{Choi2014}, the performance of 3D nano-printed actuators is comparable. A significant advantage of 3D nano-printing lies in its ability to monolithically integrate the lens during fabrication. In contrast, conventional actuators typically require post-fabrication integration of the lens through methods such as polymer droplet curing \cite{Wu2006, Chen2008}, glass reflow \cite{Yoo2012}, or pick-and-place assembly \cite{Liu2014}. These approaches either constrain design freedom for the optical element or introduce aberrations due to placement tolerances.

\section{Conclusion}
We discussed a MEMS scanner with a monolithically integrated optical lens, fabricated via two-photon polymerization. The design ensures mechanical stability with a resonant frequency larger than \SI{300}{\hertz} while maintaining a low peak power consumption of \SI{81}{\milli\watt}.
The combination of an inherently aligned lens with near diffraction-limited performance, mechanical robustness, and low power consumption is expected to enable the integration of optomechanical devices fabricated by two-photon polymerization into practical applications. To improve the precision of the system, the integration of a magnetic position sensing mechanism could enable real-time position determination for control. This approach, using a three-dimensional Hall sensor, enables accurate position tracking by mapping the magnetic field vector to the spatial position of the actuator through a one-time calibration process \cite{MagneticPositionSensing}. Such a feedback system could mitigate hysteresis and self-heating effects. In addition, the MEMS scanner could dynamically adapt to operational changes, such as fluctuations in ambient temperature, improving its performance in applications that require high precision and reliability.

\subsection*{Disclosures}

The authors declare that there are no financial interests, commercial affiliations, or other potential conflicts of interest that could have influenced the objectivity of this research or the writing of this paper.

\subsection* {Code, Data, and Materials Availability} 

Data underlying the results presented in this paper are not publicly available but may be obtained from the authors upon reasonable request.

\subsection* {Acknowledgments}

We acknowledge the support from the European Union’s Horizon 2020 research and innovation program, which provided funding for Aybuke Calikoglu’s involvement in this study. We also thank Yanis Tage for the introduction to iterative shape compensation and for providing the beam profiler.

\bibliographystyle{spiejour}
\bibliography{references}   

\appendix    

\section{Coil fill factor and optimization}

To determine the fill factor of the coils fabricated using enameled copper wire (1570225, TRU Components, Conrad Electronics SE, Germany) and a custom winding machine, we used a polished cross-section of a coil. The fill factor is \SI{70}{\percent}.

\begin{figure}[!htb]
	\begin{center}
		\begin{tabular}{c}
			\includegraphics[width=7cm]{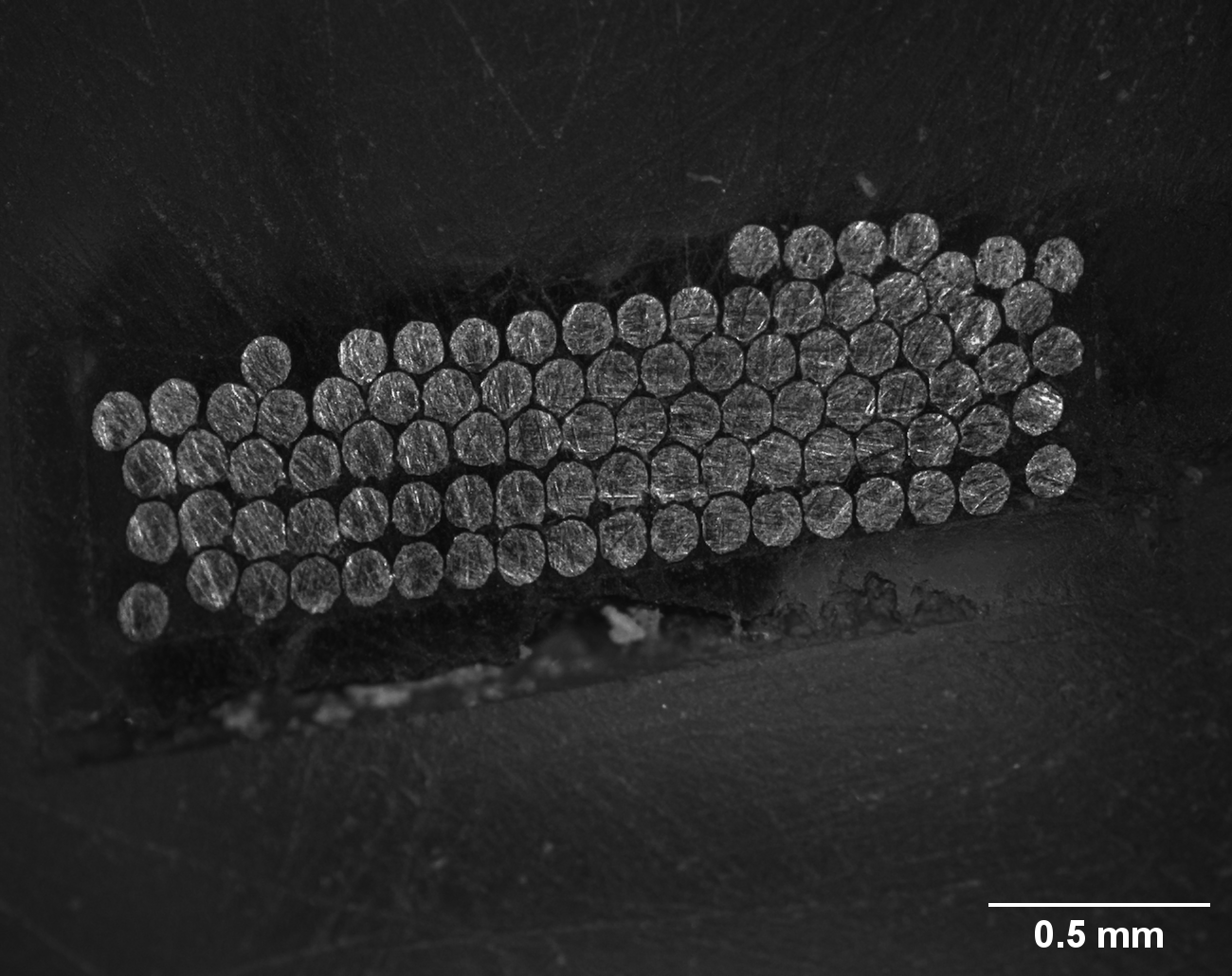}
		\end{tabular}
	\end{center}
	\caption[Cross-section of a coil]
	{ \label{fig:appendix:coilpolish}
		Polished cross-section of a coil. The fill factor is \SI{70}{\percent}.} 
\end{figure} 

\begin{figure}[!htb]
	\begin{center}
		\begin{tabular}{c}
			\includegraphics[width=14cm]{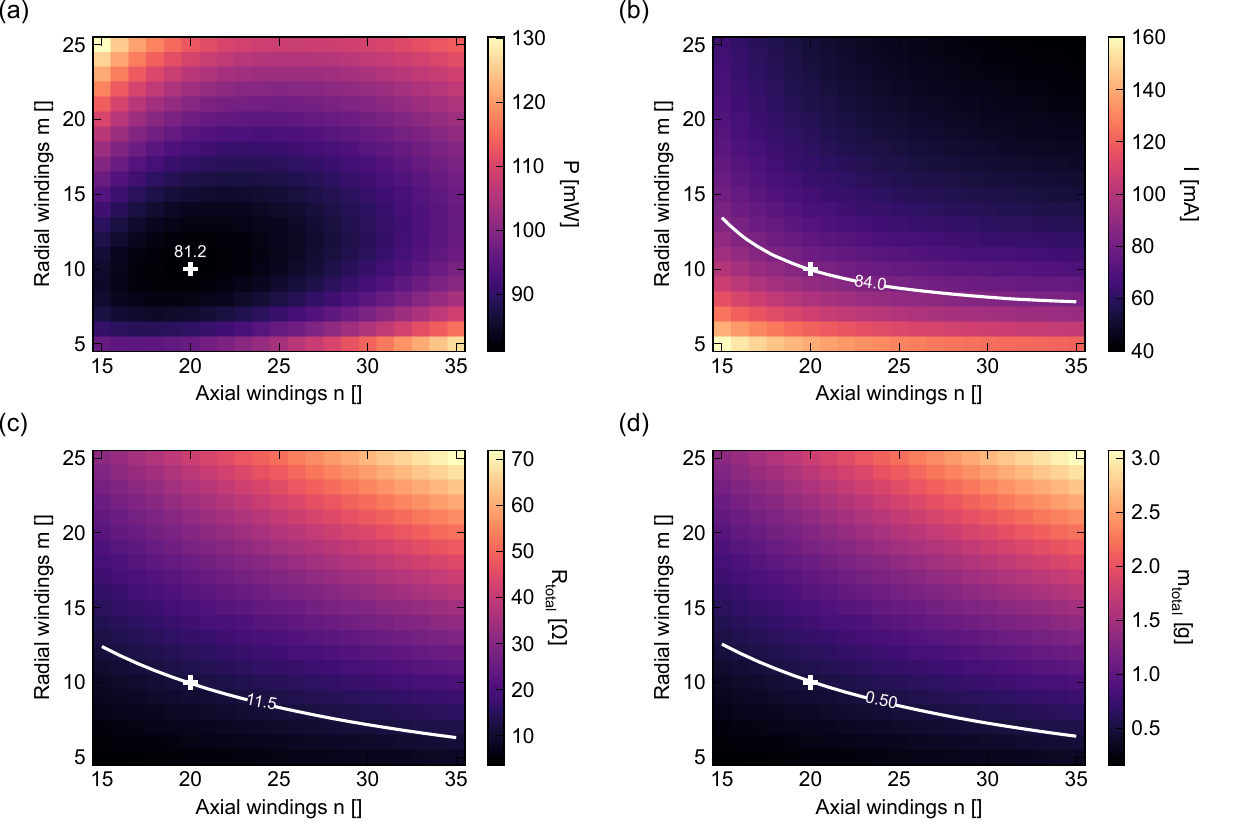}
		\end{tabular}
	\end{center}
	\caption[Coil optimization]
	{ \label{fig:appendix:coiloptimization}
		Optimization of the number of axial and radial windings of the coil pair based on the application of the Biot–Savart law to each conductor loop of the coil. The power consumption reaches a minimum of \SI{81.2}{\milli\watt} for 20 axial and 10 radial windings (a), generating a magnetic field gradient of \SI{3.03}{\tesla\per\meter} at a current of \SI{11.5}{\milli\ampere} (b). The resistance of the coil pair is 11.5~$\Omega$ (c) at a total mass of \SI{505}{\milli\gram} (d).} 
\end{figure}

Numerical calculations yield the current, resistance, mass, and power consumption of the coil, as shown in Fig.\,\ref{fig:appendix:coiloptimization}. For 20 axial and 10 radial windings, the power consumption reaches a minimum at \SI{81.2}{\milli\watt}. The current is \SI{84}{\milli\ampere} and a magnetic field gradient of \SI{3.03}{\tesla\per\meter} is generated. 

\section{Magnetic field gradient as function of radial and axial position}

\begin{figure}[ht]
	\begin{center}
		\begin{tabular}{c}
			\includegraphics[width=14cm]{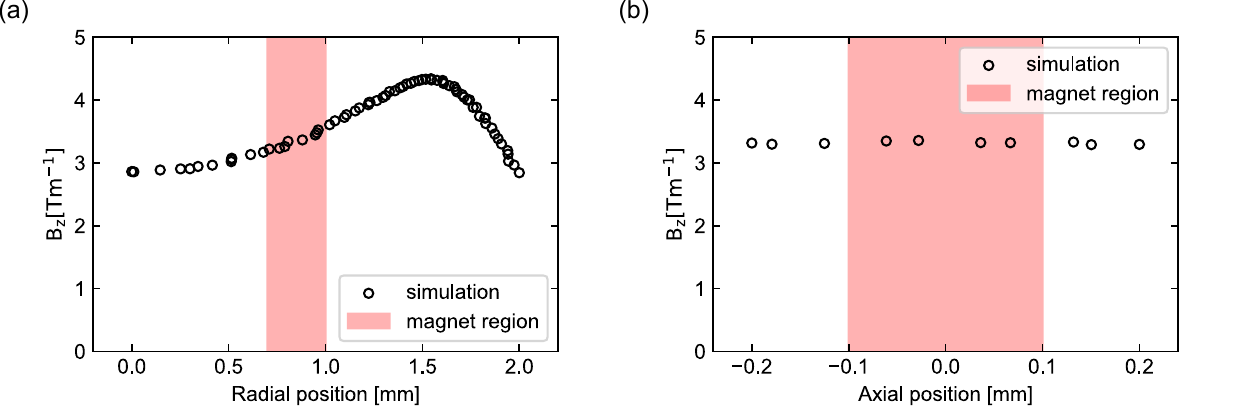}
		\end{tabular}
	\end{center}
	\caption[Magnetic field gradient in radial direction]
	{ \label{fig:appendix:gradient_simulation}
		Simulation of the z-component magnetic field gradient produced by the coils for a current of \SI{84}{\milli\ampere} shows a gradient of \SI{2.87}{\tesla\per\meter} at the center, which closely matches the calculated value of \SI{3.03}{\tesla\per\meter}. (a) In the region of the micro-magnet, the magnetic field gradient is higher than at the center. (b) The magnetic field gradient along the z-axis does not change within the actuation range. The magnetic field gradient is simulated at a radial position of \SI{0.85}{\milli\meter} (the inner radius of the magnet is \SI{0.7}{\milli\meter}, the outer radius is \SI{1}{\milli\meter}.}
\end{figure}

We simulated the magnetic field gradient using the COMSOL Multiphysics\textsuperscript{\tiny\textregistered} AC/DC module. The gradient at the center is \SI{2.87}{\tesla\per\meter}, which closely matches the calculated value of \SI{3.03}{\tesla\per\meter} obtained using the Biot–Savart law. As shown in Fig.\,\ref{fig:appendix:gradient_simulation}, the magnetic field gradient in the region of the micro-magnet is larger than at the center of the coil. This increased magnetic field gradient in the region of the magnet is not accounted for by the calculation, which assumes a uniform magnetic field gradient based on the value at the center of the coil. 

\section{Dimensions of actuator}

\begin{figure}[!htb]
	\begin{center}
		\begin{tabular}{c}
			\includegraphics[width=14cm]{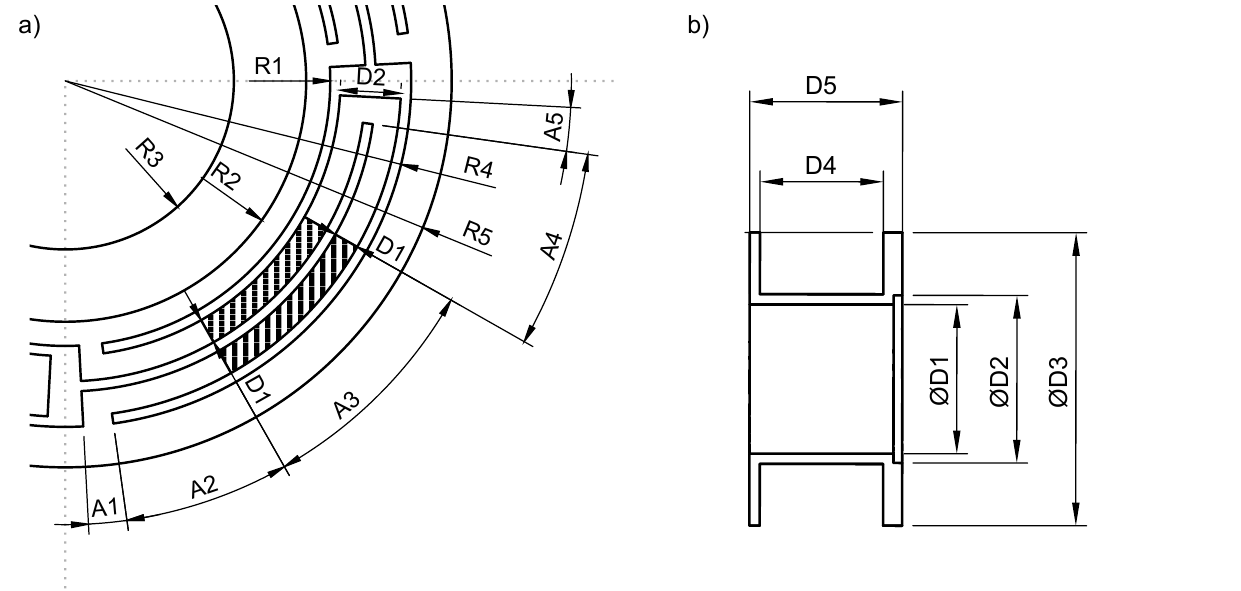}
		\end{tabular}
	\end{center}
	\caption[Technical drawing coil and actuator]
	{ \label{fig:techdraw}
		Mechanical dimensions of the springs (a) and the coil (b).} 
\end{figure}

Figure\,\ref{fig:techdraw} a presents a technical drawing, defining key dimensions of the actuator, while Fig.\,\ref{fig:techdraw}b provides a similar representation of the coil geometry. The numerical values assigned to each variable are summarized in Tab.\,\ref{tab:mech_dimensions}.

\begin{table}[!htb]
	\caption[Mechanical dimensions springs and coil]{Mechanical dimensions of the springs and the coil.} 
	\vspace{10pt}
	\label{tab:mech_dimensions}
	\centering
	\def\arraystretch{1.5}
	\setlength{\tabcolsep}{4pt}
	\begin{tabular}{lccll}
		\hline 
		\multicolumn{2}{c}{Spring} &  & \multicolumn{2}{c}{Coil holder} \\ \cline{1-2} \cline{4-5} 
		
		\multicolumn{1}{c}{Dimension} & \multicolumn{1}{c}{Value} & & \multicolumn{1}{c}{Dimension} & \multicolumn{1}{c}{Value} \\ \hline
		D1                       & \multicolumn{1}{c}{\SI{105}{\micro\meter}}   &  & D1                      & \multicolumn{1}{c}{\SI{2.90}{\milli\meter}}     \\ \cline{1-2} \cline{4-5} 
		D2                       & \multicolumn{1}{c}{\SI{252}{\micro\meter}}    & & D2                       & \multicolumn{1}{c}{\SI{3.30}{\milli\meter}}     \\ \cline{1-2} \cline{4-5}
		R1                       & \multicolumn{1}{c}{\SI{1.10}{\milli\meter}}   &  & D3                       & \multicolumn{1}{c}{\SI{5.70}{\milli\meter}}     \\ \cline{1-2} \cline{4-5}
		R2                       & \multicolumn{1}{c}{\SI{1.00}{\milli\meter}}   &  & D4                       & \multicolumn{1}{c}{\SI{2.40}{\milli\meter}}     \\ \cline{1-2} \cline{4-5}
		R3                       & \multicolumn{1}{c}{\SI{0.70}{\milli\meter}}   &  & D5                       & \multicolumn{1}{c}{\SI{2.97}{\milli\meter}}     \\ \cline{1-2} \cline{4-5}
		R4                       & \multicolumn{1}{c}{\SI{1.44}{\milli\meter}}   &  &                          &                           \\ \cline{1-2}
		R5                       & \multicolumn{1}{c}{\SI{3.21}{\milli\meter}}    & &                          &                           \\ \cline{1-2}
		A1                       & \multicolumn{1}{c}{\SI{5.0}{\degree}}     &        &                  &                           \\ \cline{1-2}
		A2                       & \multicolumn{1}{c}{\SI{21.6}{\degree}}     &     &                     &                           \\ \cline{1-2}
		A3                       & \multicolumn{1}{c}{\SI{30.8}{\degree}}     &    &                      &                           \\ \cline{1-2}
		A4                       & \multicolumn{1}{c}{\SI{21.6}{\degree}}     &   &                       &                           \\ \cline{1-2}
		A5                       & \multicolumn{1}{c}{\SI{5.0}{\degree}}     &    &                      &                        \\ \hline
		
	\end{tabular}
\end{table}

\newpage

\section{Correlation between displacement drift and coil temperature}
\label{sect:misc}

The time constants for the drift closely match the time constants of the coil temperature for currents of \SI{67.3}{\milli\ampere} and \SI{88.6}{\milli\ampere}. There is a strong linear correlation between drift and coil temperature (R\textsuperscript{2}\,=\,0.99\,/\,1.00), as depicted in Fig.\ref{fig:longterm_quasistatic}d. 

\begin{figure}[!htb]
	\begin{center}
		\begin{tabular}{c}
			\includegraphics[width=14cm]{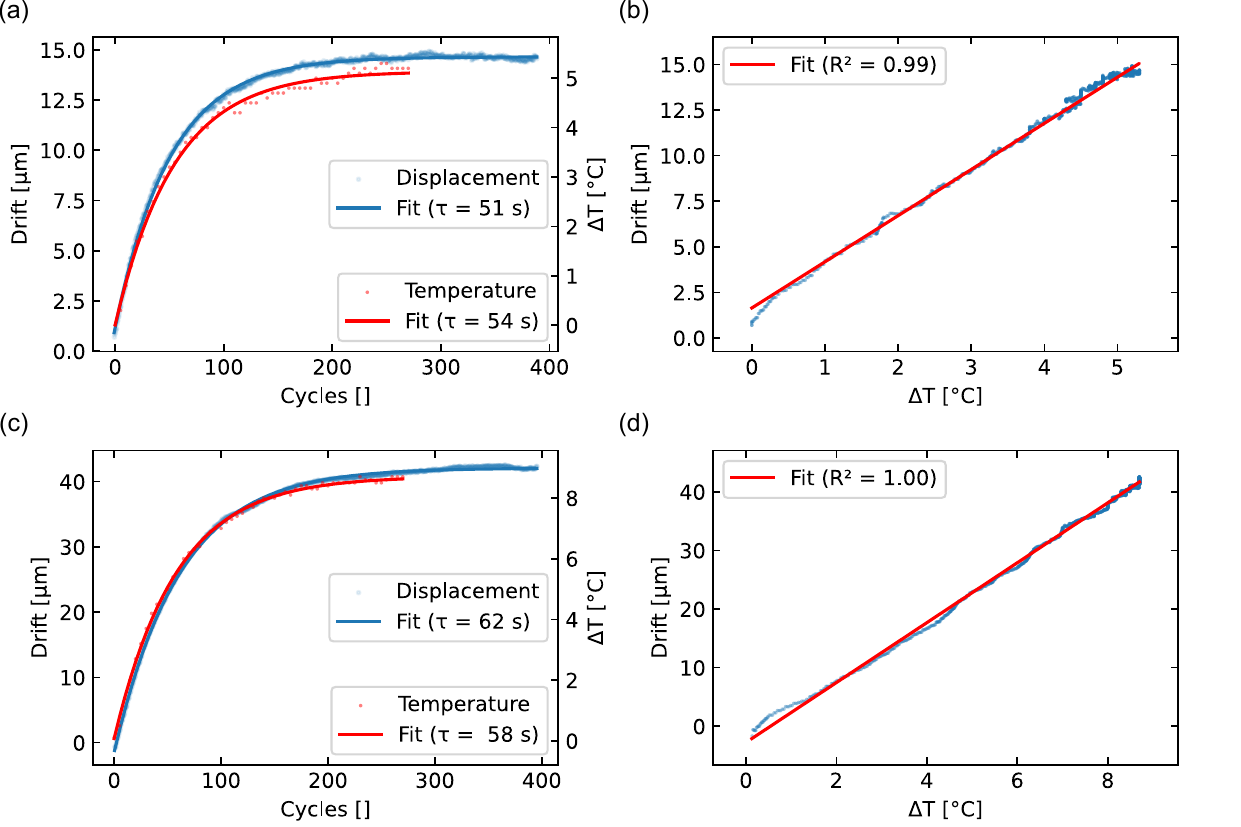}
		\end{tabular}
	\end{center}
	\caption[Extended measurement of drift and coil temperature] {\label{fig:appendix:longtermtemperature}
		Measurement of drift and coil temperature for currents of \SI{67.3}{\milli\ampere} / \SI{88.6}{\milli\ampere}. (a,c) Evaluation of displacement drift and coil temperature over time. (b,d) Linear correlation between displacement drift and coil temperature.}
\end{figure}

\end{document}